\documentstyle[12pt,aps,prc,preprint]{revtex}
\begin{document}

\begin{titlepage}
\title{\vspace*{10mm}\bf Medium effects in K$^+$
nuclear interactions \thanks{Dedicated to the memory of Carl Dover, a friend,
colleague and scholar.}}
\vspace{6pt}

\author{ E.~Friedman$^a$, A.~Gal$^a$, J.~Mare\v{s}$^b$ \\
$^a${\it Racah Institute of Physics, The Hebrew University, Jerusalem 91904,
Israel\\}
$^b${\it Nuclear Physics Institute, 25068 \v{R}e\v{z}, Czech Republic}}

\vspace{4pt}
\maketitle

\begin{abstract}
Total $(\sigma_T)$ and reaction $(\sigma_R)$ cross sections are derived
self consistently from the attenuation cross sections measured in
transmission experiments at the AGS for K$^+$ on $^6{\rm Li}$, C, Si and Ca
in the momentum range of 500-700 MeV/c by using a $V_{opt} =
t_{eff}(\rho)\rho$ optical potential. Self consistency requires, for
the KN in-medium $t$ matrix, that Im $t_{eff}(\rho)$ increases
linearly with the average nuclear density in excess of a threshold value of
0.088$\pm$0.004 fm$^{-3}$. The density dependence of Re $t_{eff}(\rho)$
is studied phenomenologically, and also applying a relativistic mean field
approach, by fitting to the $\sigma_T$ and $\sigma_R$ values. The real part
of the optical potential is found to be systematically less repulsive with
increasing energy than expected from the free-space repulsive KN interaction.
When the elastic scatttering data for $^6{\rm Li}$ and C at 715 MeV/c
are included in the analysis, a tendency of Re $V_{opt}$ to generate an
attractive pocket at the nuclear surface is observed.
\newline
$PACS$: 25.80.Nv; 13.85.Lg; 13.75.-n; 21.30.Fe
\newline

\end{abstract}\vspace{1cm}
\end{titlepage}
\section{Introduction}
\label{sec:intro}
The KN interaction for incoming momenta up to 1 GeV/c is fairly weak, with
typical total cross reaction values of 10-15 mb. It is expected then
\cite{DWa82} that K-nuclear interactions be well described, in terms of the
free-space KN $t$ matrix and the nuclear density $\rho$, by the
first-order optical potential $V_{opt} = t \rho$. However, K$^+$ nucleus
total cross section values \cite{Bug68,Mar90,Kra92,Saw93,Wei94,FGW97} and
reaction cross section values \cite{FGW97} derived from attenuation cross
sections measured in transmission experiments, as well as elastic
\cite{Mar82,Mic96} and inelastic differential cross sections
\cite{Mar82,CSP97} and quasifree spectra \cite{Kor95}, all point to a
substantial departure of
the order of 10-20\% from the predictions of a simple
$t \rho$ potential, even when many conventional nuclear medium effects
are incorporated \cite{SKG84,CEr92}. Several nonconventional medium
effects have also been proposed to remedy the failure of the
first-order optical potential approach, such as nucleon swelling \cite{SKG85},
or density dependent vector meson masses \cite{BDSW88}, or polarizing
the nuclear medium by $\bar{\rm N}$N excitations \cite{CLa92}, or
meson exchange currents \cite{JKo92,GNO95}, but as we shall discuss
in the concluding section, none of these models provides a
satisfactory solution to the discrepancy.

It was pointed out by Friedman et al. \cite{FGW97} that values of
$\sigma_R$ and $\sigma_T$ derived from transmission measurements are not model
independent. Indeed a knowledge of the Coulomb-nuclear interference
contribution at small angles is required in order to extrapolate these
measurements to zero degrees, where the total and reaction cross sections are
determined. Traditionally, an optical potential is assumed for the purpose of
extrapolation, even though it is not guaranteed that a particularly chosen
$V_{opt}$ yields by calculation the same values of $\sigma_R$ and $\sigma_T$
which eventually are derived by using it throughout the extrapolation
of the data. In this sense, most existing derivations of total cross sections in
the literature do not satisfy the basic requirement of {\it self
consistency}. The question of self consistency, posed for $V_{opt}$,
is whether or not optical potentials which are constructed empirically to fit
$\sigma_R$ and $\sigma_T$ values derived from transmission experiments
lead to the same values, within errors, when used in reanalyzing these
same experiments. Friedman, Gal and Mare\v{s} \cite{FGM97} have recently
discussed this issue, showing that self consistency for K$^+$ nucleus cross
sections can be achieved by using a density dependent optical potential
$V_{opt} = t_{eff}(\rho)\rho$ with Im $t_{eff}(\rho)$ depending linearly
on the average density in excess of a threshold value.

In this paper we report $\sigma_T$ and $\sigma_R$ values for K$^+$
interactions with $^6{\rm Li}$, C, Si and Ca, as derived self consistently
from the attenuation cross sections measured in transmission experiments
\cite{Wei94} at the AGS, at 488, 531, 656 and 714 MeV/c. Previous background
as well as a preliminary report for 714 MeV/c is given in Ref.
\cite{FGM97}. Section \ref{sec:potl} describes the wave equation and the
optical potentials used in analyzing the data and Section \ref{sec:intres}
summarizes the reaction and total cross sections obtained from a self consistent
analysis of the
transmission measurements. The phenomenology of Re $V_{opt}$ is
discussed in Section \ref{sec:realpot}, by studying its density dependence.
Some of the calculations include also fits to the angular distribution
data of Ref. \cite{Mic96}. A relativistic mean field approach to Re
$V_{opt}$ is presented in Section \ref{sec:RMF}. Both approaches suggest that
Re $V_{opt}$ becomes less repulsive with increasing energy than expected from
the free-space repulsive KN $t$ matrix interaction. Section \ref{sec:S&D}
consists of a summary and discussion.

\section{Optical potentials}\label{sec:potl}

Following Ref. \cite{FGM97}, the interaction of K$^+$ with nuclei
is described in the present work by the Klein Gordon equation

\begin{equation} \label{equ:KG1}
\left[ \nabla^2 + k^2 - (2\varepsilon^{(A)}_{red}V -
V^2_c)\right] \psi = 0~~ ~~(\hbar = c = 1)
\end{equation}

\noindent
where $k$ and $\varepsilon^{(A)}_{red}$ are the
wave number and reduced energy respectively in the c.m.
system, $(\varepsilon^{(A)}_{red})^{-1}=E_p^{-1}+E_t^{-1}$ in
terms of the c.m. energies for the projectile and target
particles, respectively. $V_c$ is the Coulomb potential due to the
charge distribution of the nucleus, and $V = V_c + V_{opt}$.
The simplest possible $t\rho$ form  for the optical potential is

\begin{equation} \label{equ:potl}
2 \varepsilon^{(A)}_{red}V_{opt}(r)=-4\pi F_k
b_0\rho(r) ,~~ ~~ F_k = \frac{M_A\sqrt s}{M(E_t+E_p)}~~,
\end{equation}

\noindent
where  $F_k$ is a kinematical factor resulting from the
transformation of amplitudes between the KN and the K$^+$ nucleus
c.m. systems and $b_0$ is the value of the KN scattering amplitude
in the forward direction. $M$ is the free nucleon mass,  $M_A$ is
the mass  of the  target nucleus, $\sqrt s$ is the total
projectile-nucleon energy in their c.m. system and the nuclear
density distribution $\rho(r)$ is normalized to $A$, the number of
nucleons in the target nucleus. It is instructive to note that,
in the eikonal approximation, the probability attenuation
produced by the potential (2) for the propagation law implied by
using the wave equation (\ref{equ:KG1}) (for $V_c = 0$) is given
by

\begin{equation} \label{equ:eik}
{{\left| {\psi _{eik}\left( {b,z} \right)} \right|^2}
\mathord{\left/ {\vphantom {{\left| {\psi _{eik}\left( {b,z}
\right)} \right|^2} {\left| {\psi _{eik}\left( {\vec b,-\infty }
\right)} \right|^2}}} \right. \kern-\nulldelimiterspace} {\left|
{\psi _{eik}\left( {b,-\infty } \right)} \right|^2}}=\exp
\left( {-\sigma \int\limits_{-\infty }^z {\rho \left( {b,\zeta }
\right)d\zeta }} \right)\  ,
\end{equation}
where $\sigma$ is the KN total cross section. Eq.(\ref{equ:eik})
corresponds to a semiclassical picture of the K$^+$ projectile
propagating forward with a mean free path $(\sigma \rho)^{-1}$
in a medium of density $\rho$, thus justifying the use of the
wave equation (\ref{equ:KG1}) with $V_{opt} = t\rho$ of Eq.(\ref{equ:potl}).
Other relativistic wave equations were
tested for comparison in Refs. \cite{FGW97,FGM97} and yielded
cross section values
practically the same within few percents as those presented
here. As for our final conclusions, details do slightly change using
other wave equations, but the overall picture remains unchanged.

A convenient `macroscopic' (MAC) parameterization of nuclear
densities is in terms of a 2- or 3- parameter Fermi function or,
for light nuclei, in terms of a modified harmonic oscillator
density. Experimental information on nuclear charge densities is
often presented using these functions \cite{VJa87}, and the proton
densities may then be obtained by properly unfolding the finite
size of the proton. Such densities are adequate around the nuclear
surface where the root-mean-square (r.m.s.) radius is determined.
However, such distributions cannot be valid far outside the
nucleus because they do not possess the correct exponential
fall-off, as determined by the binding energy of the least bound
particles. In that region it is more appropriate to use
single-particle (SP) densities or  densities obtained from a more
realistic Hartree-Fock (HF) calculation. A convenient way for generating
`microscopic'  or SP nuclear densities is to fill the appropriate
single particle  levels in a Woods-Saxon potential (separately for
protons and neutrons). The radius parameter of the potential is
then adjusted so that the resulting proton density distribution
reproduces the r.m.s. radius of the charge distribution (after
folding in the finite size of the proton) whilst at the same time
the depth of the potential is adjusted so that the binding energy
of the least bound particle is set to the corresponding separation
energy, thus ensuring the correct fall-off of the density at large
radii. Using this method, which is obviously a simplification,
rather realistic density distributions are obtained, particularly
outside the nuclear surface. The two models, namely the MAC and the
SP models, will be used here.

Fits to the data may be made by adjusting a complex parameter
$t_{eff}$  which is related to an effective forward scattering
amplitude $b_{eff}$ as follows:

\begin{equation} \label{equ:teff1}
2 \varepsilon^{(A)}_{red}t_{eff}=-4\pi F_k
b_{eff}~~.
\end{equation}
In Ref. \cite{FGM97} it was shown that very good
{\it self consistent} fits to the data
could be obtained if  the imaginary part of $t_{eff}$ was made
density dependent according to the empirical expression:

\begin{equation}\label{equ:teff}
t_{eff}(\rho)={\rm Re}~t_{eff}+i{\rm Im}~t_{eff}\left[1+\beta
(\overline{\rho}-\rho_{th})\Theta(\overline{\rho}-\rho_{th})\right]
\quad,
\end{equation}
where the average nuclear density is defined as follows:

\begin{equation} \label{equ:rhobar}
\overline{\rho}=\frac{1}{A}\int\rho^2d{\bf r}.
\end{equation}

\noindent
Values of $\overline{\rho}$ can be obtained from HF calculations
or from the simpler SP calculations described above. Results for the
four nuclei of the present work are given in Table \ref{tab:rhobar}, where
the uncertainties for C, Si and Ca are about 1\%.
It was found that the values

\begin{equation}\label{equ:beta}
\beta=(13.0\pm 3.4)~{\rm fm}^3 \quad , \qquad \rho_{th}=(0.088
\pm 0.004)~{\rm fm}^{-3} \quad ,
\end{equation}

\noindent
provide  excellent self consistent fits  at all
four momenta for which transmission
measurements on $^6{\rm Li}$, C, Si and Ca were made at the AGS
\cite{Mar90,Kra92,Saw93,Wei94}.
Attempting to avoid the use of a threshold density $\rho_{th}$
in Eq.(\ref{equ:teff}) by using powers of the average or local
density $\rho(r)$ failed to produce fits to the data. Note that
the value of the threshold density (Eq.(\ref{equ:beta})) is
considerably larger than the average density of $^6$Li, while
comfortably smaller than $\overline{\rho}$ of the other target
nuclei. This reflects the failure to reconcile the $^6$Li data
with the data for the denser nuclei, unless the specific density
dependence given by Eq.(\ref{equ:teff}) is assumed. This density
dependence, for the parameters of Eq.(\ref{equ:beta}), amounts to
20-30\% enhancement, depending on the nucleus.

Results of the self consistent
fits are given in Table \ref{tab:fits1}, essentially
taken from Ref. \cite{FGM97}. Listed are values of the fitted KN
effective forward scattering amplitude $b_{eff}$
which are determined exclusively by fitting to the $^6$Li data.
It was demonstrated in Ref. \cite{FGM97} that in the present incoming
momentum range, and for a KN interaction strength as weak as here encountered,
using an optical potential to describe scattering off as light a nucleus
as $^6$Li incurs errors of no larger than 2\%. With respect to the
free-space KN scattering amplitude $b_{0}$, the values listed for
Im~$b_{eff}$ show enhancement of 5-15\% with increasing momentum,
in quantitative agreement with the estimates of meson exchange currents
effects \cite{GNO95}.
The values listed for Re~$b_{eff}$ show a considerably larger
departure. The KN in-medium $t_{eff}$ is substantially less
repulsive than the free-space $t$. It was pointed out in Ref. \cite {FGM97}
that such a trend might be expected partly due to the proximity of the
K$^*$N and K$\Delta$ channels, but no quantitative estimates of this effect
have been made.

\section{Integral cross sections from transmission measurements}
\label{sec:intres}
As explained above and in Ref. \cite{FGM97}, we have succeeded in obtaining
self consistency in the analysis of transmission experiments by
rescaling the imaginary part of the optical potential with the
{\it average} nuclear density above a threshold value. No explanation could be
given to this empirical result, but we note that with such a rescaling,
which was found to be independent of beam momentum, we are able to obtain
reaction and total cross sections which are consistent with the optical
potential used to extract these quantities from the results of
transmission measurements. Table \ref{tab:res} summarizes the reaction
and total cross sections at the four beam momenta. The uncertainties
are the same as quoted in the earlier publication \cite{FGW97}, and they
reflect statistical errors only. Systematic uncertainties
are associated mostly with the optical  model input which is inherent
in the analysis \cite{AMa93,FGW97}. These were estimated to be as large as
5-10\%. Indeed it can be seen that the results in Table \ref{tab:res}
differ from those of Ref. \cite{FGW97} by about 5\%, always exceeding
the previous values.

\section{Phenomenology of the real potential}\label{sec:realpot}
The modification of the conventional optical potential, which made possible
the self consistent analysis and consequently led to the reasonably
good fits to the data, was the introduction of rescaling of the
imaginary potential with the factor
$1+\beta(\overline{\rho}-\rho_{th})\Theta(\overline{\rho}-\rho_{th})$.
In addition to this peculiar behaviour of the imaginary potential
it is seen from Table \ref{tab:fits1} that systematic trends can
be obtained from the real part of the best fit potentials. More
specifically, the real part of the optical potential is found to
be less repulsive than what is expected from the $t\rho$
model. The loss of repulsion seems to grow with increasing
momentum, and at the two highest momenta the real part of the
potential is essentially consistent with zero. To study further
the real part of the optical potential we introduce a
phenomenological density dependence (DD) also into the real
potential such that the real part of $t_{eff}$ is made to depend
on the nuclear density as follows:

\begin{equation} \label{DDdef}
b_{eff} \longrightarrow b_{eff} +
B_0\left[{{\rho(r)}\over{\rho(0)}}\right]^\alpha.
\end{equation}
This form of a potential was found to be quite successful in studies
of DD effects in kaonic atoms \cite{FGB93} and in
$\Sigma^-$ atoms \cite{BFG94,MFG95}. It must be emphasized that
this is a purely phenomenological form  and that no significance is
attached to e.g.  the exponent $\alpha$. However, when $b_{eff}$ is
constrained to have the corresponding value $b_0$ for the
free-space KN interaction and when $\alpha>$0, then the low
density limit \cite{DHL71}  is respected by the real potential. Using
such a DD potential for the real part and using the same
form, Eq.(\ref{equ:teff}), for the imaginary
potential  (varying only Im $b_{eff}$), we performed
$\chi^2$ fits, separately at each energy,  to the reaction and
total cross sections of Table \ref{tab:res}. In the left hand side
of Table \ref{tab:fits2} are given the parameters of the real potential
from unconstrained fits, where Re $b_{eff}$, Im $b_{eff}$, Re $B_0$
and $\alpha$ were varied. Fits of similar quality  could be
obtained for a wide range of values for $\alpha$ and the results
shown here are for $\alpha = 2$. The errors quoted are obtained
from the $\chi^2$ fit procedure and in this case there are strong
correlations between Re $b_{eff}$ and Re $B_0$. Some decrease in
the values of $\chi^2$ compared to the corresponding values in
Table \ref{tab:fits1} is evident. In the right hand side of Table
\ref{tab:fits2} are given the results of constrained fits, as
explained above. This time the value of $\alpha$, which again was
held fixed during the fits, was 0.5 as it gave slightly better
results. In this case there are no correlation effects and the
errors of Re $B_0$ apply to the sum of Re $b_0$ and Re $B_0$.
This sum can be compared directly to the values of the
density independent parameter Re $b_{eff}$ of Table
\ref{tab:fits1}. It is seen that the two sets of numbers agree
remarkably well, thus supporting the preliminary conclusions
from Table \ref{tab:fits1}
regarding the real part of the potential.

Comparing the results of the two kinds of fits in Table \ref{tab:fits2}, there
is a distinct difference regarding the behaviour of the real potential
at large radii. Whereas the constrained fit is dominated by the
free-space KN interaction at large radii and consequently is
repulsive at all four momenta of the present study, the
unconstrained fit leads nominally to an attractive potential at
large radii for 656 and 714 MeV/c. A way of bypassing the problem
of correlations between errors is to apply a `notch test' method
\cite{Fri95} to estimate the uncertainties in the real potential as
a function of position. Figure \ref{fig:int} shows the real part
of the optical potential for 714 MeV/c K$^+$ on carbon as obtained
from the unconstrained fits, together with the errors from the
notch test. Obviously more data is needed in order to study the
real potential.

Differential cross sections for the elastic scattering of K$^+$ by $^6$Li
and C are available at the highest momentum covered by the transmission
measurements \cite{Mic96} and it is therefore possible to include them
in the optical model fits.
The adequacy of the optical potential Eq.(\ref{equ:potl}) was tested
by comparing angular distributions calculated for this
potential with angular distributions calculated for an equivalent
potential \cite{FGW97} containing an explicit p-wave
term in the KN interaction.
Preliminary fits showed that if only
differential cross sections are used, then reasonably good fits are
possible but, however, the calculated reaction and total cross
sections are smaller than the measured values by 30 to 50\%. We
have therefore used an extended input data set that included the
differential cross sections for $^6$Li and C together with the
eight integral cross sections for $^6$Li, C, Si and Ca at 714
MeV/c. As with the fits to only the integral cross sections, in
these combined fits we have tried three different potentials:  (i)
a $t_{eff}\rho$ potential; (ii) an unconstrained DD potential;
 (iii) a constrained DD potential. In all
cases the imaginary part included the rescaling discussed above.
The normalization of the differential data was allowed to vary too, within the
quoted range of $\pm$15\% \cite{Mic96}. The first two potentials
produced good fits to the integral cross sections but only the
unconstrained DD potential led, at the same time, also to a
reasonably good fit to the differential data. The third potential,
namely, the constrained DD potential that respects the low density
limit, failed badly to reproduce simultaneously both types of data
and also required an unacceptably large renormalization of the
differential cross sections. In order to check the sensitivity of
the results to the particular nuclear densities used in the optical
potentials, the whole procedure was followed with
both MAC and SP densities, as
described above. The conclusions regarding the quality of the
various fits remained unchanged.

Examples of the resulting real potentials for C are shown in Fig.
\ref{fig:intdiff}. The continuous curves are for the unconstrained
 DD potential and the dashed curves are for the $t_{eff}\rho$
 potential. Examples for the uncertainties, as obtained from
the notch test mentioned above, are plotted for the unconstrained
DD potential based on the MAC densities. These should represent also
the uncertainties in the other potentials which are included in
the figure. Also shown are  the potentials obtained when the MAC
densities are replaced by the SP densities and it is seen that
although there are differences in details, the overall picture is
the same for the two models. If, in the constrained fits, the
parameter $\alpha$ is allowed to become negative, then better
quality fits are obtained but the low density limit is no longer
respected. The attractive 'pocket' near the nuclear surface, as
seen in Fig. \ref{fig:intdiff}, persists. Comparing this figure
with Fig. \ref{fig:int} it is seen that the real potential is a
little better determined when the differential cross sections are
included in the analysis. It is therefore concluded that the
empirical potentials fail to respect the low density limit, at
least within the present parameterization. It is doubtful whether
 a more elaborate phenomenological real potential
can be meaningfully derived from the present data.

\section{relativistic mean field potentials}\label{sec:RMF}
The Relativistic Mean Field (RMF) theory, treating nucleons as
Dirac particles interacting via (large) scalar and vector
fields, proved to be a valuable tool to describe nuclear structure
and dynamics \cite{SWa86}. Extensions of the nuclear RMF theory to
include hyperons are reviewed in Ref. \cite{MJe95}. In particular,
the RMF approach has been applied to
constructing the optical potential for elastic scattering of
$\Lambda$ and $\Sigma$ hyperons on nuclei \cite{CJM94}.
Therefore, the application of the RMF theory to K$^+$ --
nucleus scattering seems to be a natural extension and, also,
an interesting alternative to the phenomenological DD analysis.
Furthermore, fitting the RMF potentials to K$^+$ -- nucleus scattering
data has the potential of providing information on the coupling of kaons
to the meson fields involved, which is relevant for studying
the behaviour of kaons in nuclear matter, particularly in connection
to kaon condensation, and also for determining the equation of state
of strange baryonic matter present in neutron stars \cite{SMi96}.

In the RMF calculations reported below we used the common Lagrangian
density in the nucleonic sector \cite{SWa86}, with the linear
parameterization of Horowitz and Serot \cite{HSe81}.
For completeness, we also utilized the nonlinear model of Sharma et
al. \cite {SNR93}. Kaons were incorporated into the RMF model by
using kaon-nucleon interactions motivated by one boson exchange
models.
The simplest relevant form of the kaon-meson Lagrangian reads:
\begin{eqnarray}
{\cal L}_{K} = \partial_{\mu}\overline{\psi}\partial^{\mu}\psi -
m^2_K\overline{\psi}\psi
- g_{\sigma K }m_K\overline{\psi}\psi\sigma &
\nonumber \\
 - ig_{\omega K}(\overline{\psi}\partial_{\mu}\psi {\omega}^{\mu} -
\psi \partial_{\mu}
\overline{\psi}{\omega}^{\mu})
- ig_{\rho K}(\overline{\psi}{\vec{\tau}}\partial_{\mu}\psi {\vec{\rho}}^{\mu}
-&
\psi {\vec{\tau}}\partial_{\mu}
\overline{\psi}{\vec{\rho}}^{\mu}),
\end{eqnarray}
describing the interactions of kaons ($\psi$) with the scalar ($\sigma$)
and vector ($\omega$ and $\rho$) fields. As was pointed out by Schaffner and
Mishustin \cite{SMi96}, this simple form has to be extended by
the additional term
\begin{equation}\label{equ:LV2}
{\cal L}_{V^2}=(g_{\omega K}{\omega}_{\mu}+
g_{\rho K}{\vec{\tau}}{\vec{\rho}}_{\mu})^2
\overline{\psi}\psi
\end{equation}
in order to satisfy in the medium the Ward identity requiring
coupling of the vector field to a conserved current.
For isospin saturated nuclei,
the equation of motion for the kaons reduces to:
\begin{equation}\label{equ:emk}
[\partial_{\mu}\partial^{\mu} + m^2_K + g_{\sigma K }m_K
\sigma + 2i g_{\omega K}\omega_0 \partial^0 - (g_{\omega K}\omega_0)^2]
\psi = 0 \;\;\; ,
\end{equation}
where $\omega_0$ denotes the time component of the isoscalar vector field
$\omega$. It should be noted that the last (quadratic) term corresponding
to  ${\cal L}_{V^2}$ of Eq.(\ref{equ:LV2}) was omitted in the
analysis of Section \ref{sec:realpot}.
We have checked the effects of the quadratic term ($\sim\omega_0^2$),
as well as a possible extension by a term proportional to
$\sigma^2$, and found these to be negligible. It is also possible to
formulate the RMF calculations by reducing the second-order KG equation
(\ref{equ:emk}) into a Kemmer-Duffin-Petiau set of first-order coupled
equations, as was done in Ref. \cite{CHK85}.

The equation of motion can be expressed in the form of the
KG equation (Eq.(\ref{equ:KG1})) with the real part of the potential given by:
\begin{equation}\label{equ:VOP1}
{\rm Re }\;V_{opt}={{m_K}\over{2\varepsilon^{(A)}_{red}}}S +
{{E_p}\over{\varepsilon^{(A)}_{red}}}V \;\;\; ,
\end{equation}
where $S= g_{\sigma K}\sigma$ and $V= g_{\omega K} \omega_0$.
For the coupling constants $g_{\sigma K}$ and
$g_{\omega K}$ we used the constituent quark model values \cite{SMi96,BRh96},
namely
\begin{equation}\label{equ:quark}
{{g_{\sigma K}} \over {g_{\sigma N}}} = {{g_{\omega
K}}\over{g_{\omega N}}} = {1 \over 3} \;\;\; .
\end{equation}
\noindent
In order to account for recoil effects we employed, following
Ref.\cite{CHC93}, the recoil corrections given by Cooper and
Jennings \cite{CJe88}, namely, we multiplied the scalar and
vector potentials by the factors
\begin{equation}\label{equ:recoil}
R_S=\frac{M_A}{E_p+E_t},\;\;\; R_V=\frac{E_t}{E_p+E_t}.
\end{equation}
\noindent
Finally we introduced extra scaling coefficients for the scalar
($C_S$) and vector ($C_V$) potentials to be obtained from fits
to the data. Using these scaling factors we aimed at tracing
the energy dependence of the kaon couplings.
After multiplying the S and V potentials in Eq.(\ref{equ:VOP1})
by $R_S$ and $R_V$,
and by $C_S$ and $C_V$, respectively, we arrive at the expression for
the real part of the optical potential $V_{opt}$ to be used as a real potential
in Eq.(\ref{equ:KG1}):
\begin{equation}\label{equ:ReV}
{\rm Re }\; V_{opt} = \frac{m_K}{2E_p}\frac{M_A}{E_t}C_SS+C_VV .
\end{equation}
\noindent
The imaginary part of $V_{opt}$ was again modified by
letting Im $t_{eff}(\rho$)
depend linearly on the average density in excess of a threshold value
(see Eq.(\ref{equ:teff})),
with the same values of parameters $\beta$ and $\rho_{th}$ as in the
case of the phenomenological potentials of Section II.

The fits of the RMF potentials were made by gridding on values of
$C_V$ and
varying the coefficient $C_S$ together with Im~$b_{eff}$.
 This procedure was chosen because of the strong
correlation between $C_S$ and $C_V$. Searches stopped
once reasonably low values of $\chi^2$
(in our case $\chi^2<10$) were obtained.
The results for the linear RMF parameterization \cite{HSe81}
obtained from fits to the integral cross sections are summarized
in Table \ref{tab:rmffits}. It is seen that
the scaling coefficients $C_S$ and $C_V$ decrease with increasing
momentum, thus reducing significantly the contributions of both
the scalar and vector potentials. Consequently,
the resulting real part of $V_{opt}$ is essentially consistent with zero at the
highest momenta. Practically identical results (within the indicated errors)
were obtained for the nonlinear RMF parameterization. When the differential
cross sections are also included (only at 714 MeV/c),
then  the RMF  best fits
are significantly poorer than those obtained with the unconstrained DD
potentials. The resulting RMF real potential does not have any attractive
region, in contrast to the DD real potential.
The fits based on the RMF approach thus confirm the empirical results
concerning the real potential and in particular  the loss of
repulsion with increasing momentum. Note that this is opposite to the trend
of Dirac-phenomenological nucleon optical potentials which become less
attractive, even repulsive, with increasing energy \cite{CHC93}.

\section{SUMMARY AND DISCUSSION}\label{sec:S&D}
In the present work
we have derived {\it self consistently} integral cross
sections for K$^+$-nucleus interactions  from transmission experiments
\cite{Mar90,Kra92,Saw93,Wei94},  updating the procedure of Ref.\cite{FGW97}.
 These reaction and total cross sections provide 32 data points
for the study of medium effects in K$^+$ nuclear interactions. It had been
realized for the total cross sections data subset
\cite{SKG84,CEr92,SKG85,BDSW88,CLa92}, and quite recently also for the reaction
cross sections data subset \cite{FGW97,FGM97}, that the values of these
integral cross sections, for the relatively dense C, Si and Ca nuclei, exceed
substantially the predictions of the first-order optical potential $V_{opt} =
t \rho$, by up to about 25\%. This is inconceivable \cite{CEr92} for as weakly
interacting hadron as the K meson is. The free-space $t$ matrix used in these
theoretical studies, particularly its imaginary part which is proportional to
the KN total cross section $\sigma$, is usually consistent with the total cross
sections derived for the deuteron
in the same transmission experiments \cite{Wei94,FGW97}.

In the eikonal approximation (cf. Ref. \cite{CEr92} for testing its validity at
these energies), the total {\it reaction} cross section is given by

\begin{equation} \label{equ:sigR}
\sigma _R=\int {d^2b\left( {1-\exp \left( {-{2 \over {\hbar
v}}\int\limits_{-\infty }^\infty  {W\left( r \right)dz}} \right)} \right)}\  ,
\end{equation}

\noindent
where $W(r)=-$ Im $V_{opt}(r)$. It is clear that, in order to increase the
calculated value of $\sigma_R$, it is necessary to increase correspondingly
$W(r)$ beyond the $t\rho$ expression $W_{opt}(r)= \hbar v \sigma \rho(r)/2$.
Whereas a moderate increase of $W(r)$ (cf. the values of Im $b_{eff}$ listed in
Table \ref{tab:fits1} vs. the values Im $b_0$ listed there in parentheses) was
found to satisfactorily reproduce the integral cross sections for $^6$Li, no
simple extension in terms of powers of the density $\rho$,
or of the average density $\bar \rho$, could be
found to reproduce simultaneously
the integral data for the other, considerably denser nuclei. The only way found
to reproduce self consistently all the integral cross sections was by
introducing the density dependence of Eq. (\ref{equ:teff}) according to which
the (energy dependent) coefficient Im $t_{eff}$ increases linearly with
($\bar \rho - \rho_{th}$) once $\bar \rho > \rho_{th}$,
for a value of $\rho_{th}$ given by Eq. (\ref{equ:beta}) which is essentially
independent of energy. Since, by
fitting to the data, $\rho_{th}$ turned out to be intermediate between $\bar
\rho$ ($^6$Li) and $\bar\rho$ for
the other heavier nuclei (cf. Table \ref{tab:rhobar}), using
this ad-hoc prescription for Im $V_{opt}$
it is possible to bring the calculated
C, Si and Ca integral cross sections into line with the $^6$Li cross
sections. It remains an open problem to understand the significance of this
dependence and of this
fitted value of $\rho_{th}$ in terms of nonconventional reactive channels which
might open up for $\bar \rho > \rho_{th}$.

For Re $V_{opt}$ it was not  necessary to introduce as complex density
dependence as discussed above for Im $V_{opt}$. In fact, Re $t_{eff}$ could be
kept independent of density, becoming progressively less repulsive with energy,
so that Re $V_{opt} \approx 0$ for $p \sim 700$ MeV/c. It was also possible to
respect the low density limit, with a repulsive $t\rho$ term of  about
25 MeV, by adding an attractive term depending on a higher power of the
density such that the
overall Re $V_{opt}$ again becomes close to zero for $p \sim 700$ MeV/c. This
same tendency of Re $V_{opt}$ to become less repulsive with energy is also
obtained when fits are based on
 RMF scalar and vector potentials taken from structure calculations
or from nucleon-nucleus Dirac phenomenology. The RMF
scaling factors $C_S$ and $C_V$ of Eq.(\ref{equ:ReV})
become vanishingly small with increasing
energy, so that the (repulsive) vector and (attractive) scalar contributions at
$p = 714$ MeV/c are individually small, of order 5 MeV for C, with a
net repulsion given by their difference of about 5 MeV. It is
interesting to note that for K$^-$ nucleus scattering where the vector
potential has the opposite sign,
so that both $S$ and $V$ are then attractive, we would
predict an attractive potential depth of about 10 MeV. Indeed, optical model
fits \cite{ABD83} to K$^-$ $^{12}$C elastic scattering at $p = 800$ MeV/c
\cite{Mar82} yield potential depths about $30 \pm 6$ MeV, depending on the
geometry assumed for Re $V_{opt}$. This relatively weak attraction for $\bar
{\rm K}$ at intermediate energies is in stark contrast to the
attraction of order 200 MeV felt by $\bar {\rm K}$ mesons at low energies,
as derived by fitting K$^-$ atomic data \cite{FGB93}.

We have also studied differential cross sections
for elastic scattering of K$^+$
recently reported \cite{Mic96} for $^6$Li and C at $p= 715$ MeV/c.
Whereas these
data definitely require a significant amount of density dependence beyond the
$t\rho$ approximation in order to achieve reasonable good fits,
the resultant potentials are
incompatible with potentials obtained from fits to
integral cross sections only. When
fits to the combined data set at 714 MeV/c consisting of both differential and
integral cross sections are made, Re $V_{opt}$ is weakly repulsive within the
nucleus, turning into an attractive pocket at the surface which, however,
violates the low density limit. As for the fitted imaginary part of the optical
potential, the specific density dependence of Eq. (\ref{equ:teff}) is upheld
also by this extended set of data.

A brief discussion of several theoretical works is in order. Siegel et al.
\cite{SKG85} suggested that nucleon ``swelling" in the medium primarily affects
the dominant $S_{11}$ KN  phase shift by
increasing it in the nuclear medium. In the momentum range 500-700
MeV/c this amounts to increasing Im $b_0$, as required by the data, but {\it
decreasing} (the negative) Re $b_0$ contrary to the trend suggested by fitting
to the data, of increasing it to become less repulsive. A similar remark holds
against  the mechanism of dropping
vector-meson masses in the nuclear medium as suggested by Brown et al.
\cite{BDSW88}. The meson exchange current effect considered in Refs.
\cite{JKo92,GNO95} is capable of producing the increment necessary for Im $b_0$
in order to fit the integral cross sections as function of energy, but it
suggests no significant change in Re $b_0$. Finally, the most recent work by
Caillon and Labarsouque \cite{CLa92}, who consider medium effects for the
mesons exchanged between the K$^+$ and the bound nucleons, produces {\it both}
modifications required in order to fit the data, namely increasing Im
$b_0$ and Re $b_0$ simultaneously to an extent which is comparable at 700 MeV/c
with the values of $b_{eff}$ shown in our Table \ref{tab:fits1}. However, these
authors as well as  other works are  unable to produce the major medium
effect which in our phenomenological approach is expressed by modifying Im
$b_{eff}$ by the $[1 + \beta (\bar \rho - \rho_{th}) \Theta (\bar \rho -
\rho_{th})]$ factor. In this sense, no satisfactory theoretical approach yet
exists to describe K$^+$ nucleus interaction at intermediate energies. We point
out two features which make it even harder for theory to explain the integral
cross sections data: (i) the self consistent values of $\sigma_T$ published
here are always larger, by about 5\% than the previous values \cite{Wei94} used
in some of the theoretical works; (ii) there exist now $\sigma_R$ data,
published here, which indicate a similar problem for theory as the $\sigma_T$
data do. This appears to suggest that the theoretically missing
part of  the cross section
belongs to some in-medium
major reaction channels mistreated by the $t\rho$ optical model
approach and its conventional, relatively minor modifications; or that at
present theory misses some nonconventional in-medium effects that would
strongly invalidate the $t\rho$ starting point.

New inelastic data for $^6$Li and C at 635 and 715 MeV/c are being reported
\cite{CSP97}. These inelastic excitations also require to be enhanced
theoretically in order to fit the new data. Even though the corresponding
transition densities peak at the nuclear surface where $\rho < \rho_{th}$, one
cannot simply argue that these inelastic processes need not be strongly
renormalized by the medium effect specified here; ours is a {\it global}, not a
local prescription, since it involves $\bar \rho$;  not $\rho$. Of course,
more work is needed to decide whether or not these inelastic cross sections
shed new light on the problem of medium effects in K$^+$ nucleus interactions.

\vspace{15mm}
This research was partially supported by the Israel Science Foundation
(E.F.), by the U.S.-Israel Binational Science Foundation (A.G.) and by
the Grant Agency of the Czech Republic (J.M., grant No. 2020442).
A.G. and E.F. acknowledge the hospitality of the Nuclear Physics Institute
at \v{R}e\v{z} and J.M. acknowledges the hospitality of the Hebrew
University.

\begin{figure}
\caption{Real part for the K$^+$ carbon potential at 714 MeV/c
from unconstrained DD fits to integral cross sections
using MAC densities. The
error bars are determined from  notch tests (see text).}
\label{fig:int}
\end{figure}

\begin{figure}
\caption{Real part for the K$^+$ carbon potential at 714 MeV/c
from fits to integral and differential cross sections. Continuous lines
are for unconstrained DD potentials and dashed lines are for $t_{eff}\rho$
potentials. MAC and SP refer to the model used to calculate nuclear densities.}
\label{fig:intdiff}
\end{figure}

\begin{table}
\caption{Values of $\overline{\rho}\;$ (in units of fm$^{-3}$)
obtained from SP calculations.} \label{tab:rhobar}
\begin{tabular}{cccc}
$^6$Li&C&Si&Ca \\ \hline
0.049&0.103&0.110&0.107 \\
\end{tabular}
\end{table}

\begin{table}
\caption{Fits to K$^+$ nucleus $\sigma_R$ and $\sigma_T$ values
obtained by rescaling Im $V_{opt}$ by the factor
$1+\beta(\overline{\rho}-\rho_{th})\Theta
(\overline{\rho}-\rho_{th})$ with $\rho_{th}=0.088$ fm$^{-3}$,
$\beta=13.0$ fm$^3$.
Values in parentheses are for the free-space KN interaction
parameter $b_0$. Values of $\chi^2$ refer to the whole data base
of 8 points at each momentum.}
 \label{tab:fits1}
\begin{tabular}{cccc}
$p\;$(MeV/c)&Re $b_{eff}$(fm)&Im $b_{eff}$(fm)&
$\chi^2$ \\ \hline
488&$-0.154\pm$0.012&0.160$\pm$0.002&0.5\\
 &($-0.178$)&(0.153)&  \\
531&$-0.119\pm$0.012&0.186$\pm$0.002&10.3\\
 &($-0.172$)&(0.170) & \\
656&$-0.035\pm$0.062&0.241$\pm$0.002&2.1\\
 &($-0.165$)&(0.213)&   \\
714&$-0.044\pm$0.064&0.265$\pm$0.001&7.8\\
 &($-0.161$)&(0.228)&   \\
\end{tabular}
\end{table}

\newpage
\begin{table}
\caption{Reaction and total cross sections (in
mb) for K$^+$ interaction with various nuclei from self consistent
analysis of transmission measurements.}
\label{tab:res}
\begin{tabular}{ccccccccc}
\multicolumn{1}{c}{} &
\multicolumn{4}{c}{reaction} &
\multicolumn{4}{c}{total} \\  
$p\:$(MeV/c)
& $^6{\rm Li}$ & C &  Si & Ca & $^6{\rm Li}$ &
C & Si & Ca \\
\hline  488 & 67.8 &
128.4 & 276.2 & 362.5 & 77.5 & 165.4 & 373.7 & 503.2 \\
& $\pm$ 1.3 & $\pm$ 2.3 &  $\pm$ 5.1 & $\pm$
7.7 & $\pm$ 1.1 & $\pm$ 1.9 & $\pm$ 4.8 & $\pm$
7.7  \\
531 & 73.2 &
136.8 & 299.1 & 384.0 & 80.7 & 168.9 & 391.7 & 521.6 \\
& $\pm$ 0.8 & $\pm$ 1.4 &  $\pm$ 3.4 & $\pm$
4.5 & $\pm$ 0.7 & $\pm$ 1.3 & $\pm$ 3.3 & $\pm$
4.4 \\
656 & 79.0 &
148.2 & 311.8 & 408.6 & 86.4 & 179.5 & 403.2 &
548.8  \\
& $\pm$ 1.1 & $\pm$ 1.5 &  $\pm$ 3.4 &
$\pm$ 5.0 & $\pm$ 0.7 & $\pm$ 0.8 & $\pm$ 2.7 & $\pm$
4.2  \\
714 & 82.2 &
152.8 & 320.2 & 417.1 & 88.5 & 183.8 & 411.3 &
550.4 \\
& $\pm$ 1.2 & $\pm$ 1.5 &  $\pm$ 3.6 &
$\pm$ 5.5 & $\pm$ 0.6 & $\pm$ 0.9 & $\pm$ 2.3 & $\pm$
2.8 \\
\end{tabular}
\end{table}

\begin{table}
\caption{Re $V_{opt}$ parameters (in fm) from density dependent
fits to integral cross sections using MAC densities.
Constrained fits are made
to respect the low density limit.}
\label{tab:fits2}
\begin{tabular}{ccccccc} 
\multicolumn{1}{c}{ } &\multicolumn{3}{c}{unconstrained}
&\multicolumn{3}{c}
{constrained} \\
$p$ (MeV/c) &Re $b_{eff}$&Re $B_0$&$\chi^2$&Re $B_0$&Re
$B_0$+ Re $b_0$& $\chi^2$ \\ \hline
488 & $-0.175\pm$0.034& 0.033$\pm$0.054 & 0.2 & 0.029$\pm$0.014
& $-0.149\pm$0.014&0.3 \\
531 & $-0.093\pm$0.049& $-0.045\pm$0.077 & 9.9& 0.071$\pm$0.018
& $-0.101\pm$0.018& 11.2 \\
656 & 0.077$\pm$0.068&$-0.084\pm$0.080 & 1.4 & 0.146$\pm$0.052
& $-0.019\pm$0.052& 2.2 \\
714 & 0.046$\pm$0.087& $-0.130\pm$0.152 & 5.4 & 0.129$\pm$0.059
& $-0.032\pm$0.059
& 10.4 \\ 
\end{tabular}
\end{table}

\begin{table}
\caption{Coefficients of the scalar ($C_S$) and vector ($C_V$)
RMF potentials obtained from fits to the integral data. Also
shown are the coefficient of the imaginary potential and the
$\chi^2$ of the fits.}\label{tab:rmffits}
\begin{tabular}{ccccc} 
$p\;$(MeV/c)&$C_S$&$C_V$&Im $b_{eff}$(fm)&$\chi^2$ \\ \hline
488&0.66$\pm$0.10&0.40$\pm$0.05&0.150$\pm$0.002&9.3 \\
531&0.08$\pm$0.04&0.10$\pm$0.05&0.185$\pm$0.002&9.0 \\
656&0.27$\pm$0.10&0.15$\pm$0.03&0.232$\pm$0.003&8.1 \\
714&0.03$\pm$0.08&0.05$\pm$0.02&0.264$\pm$0.003&9.2 \\
\end{tabular}
\end{table}

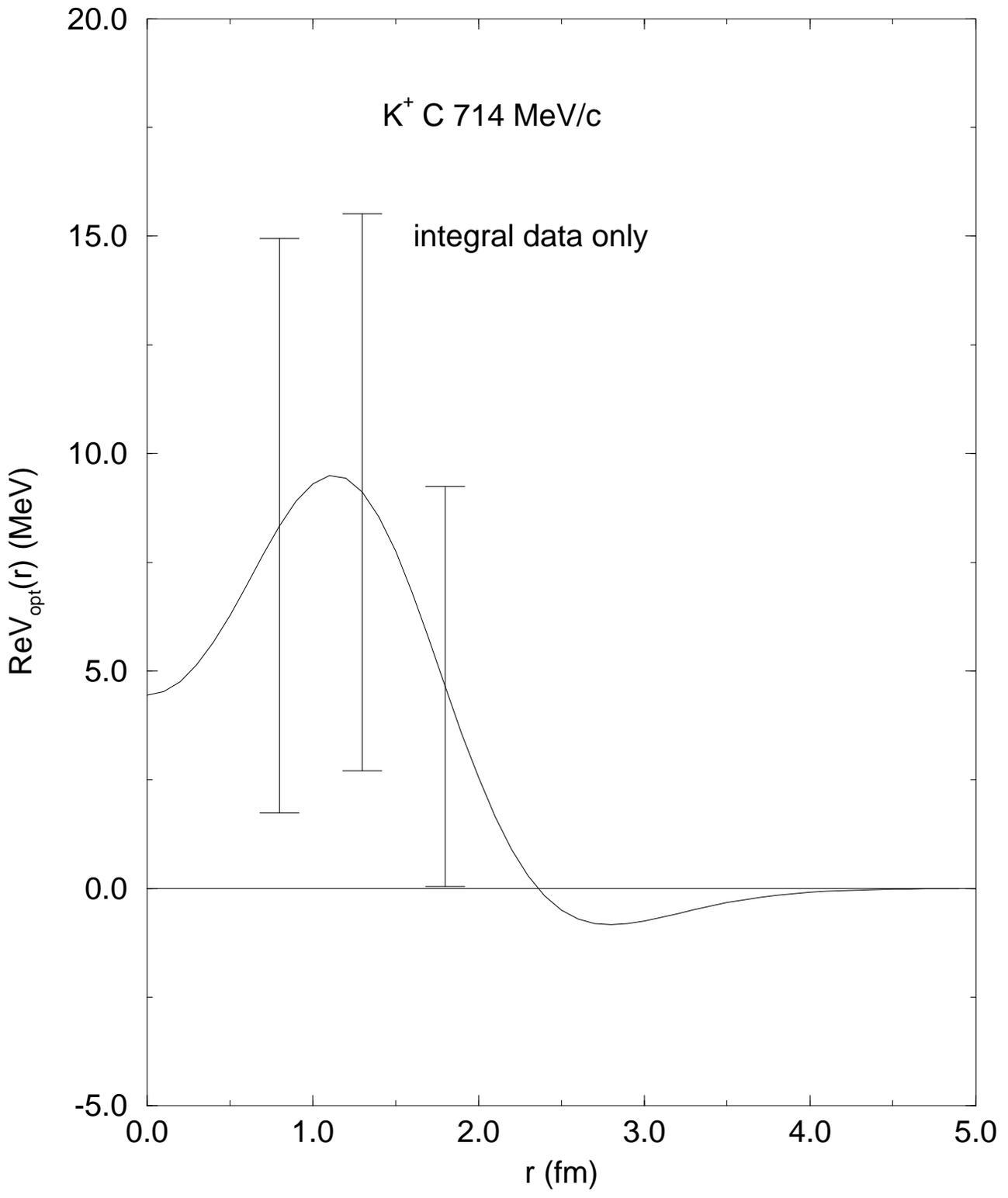
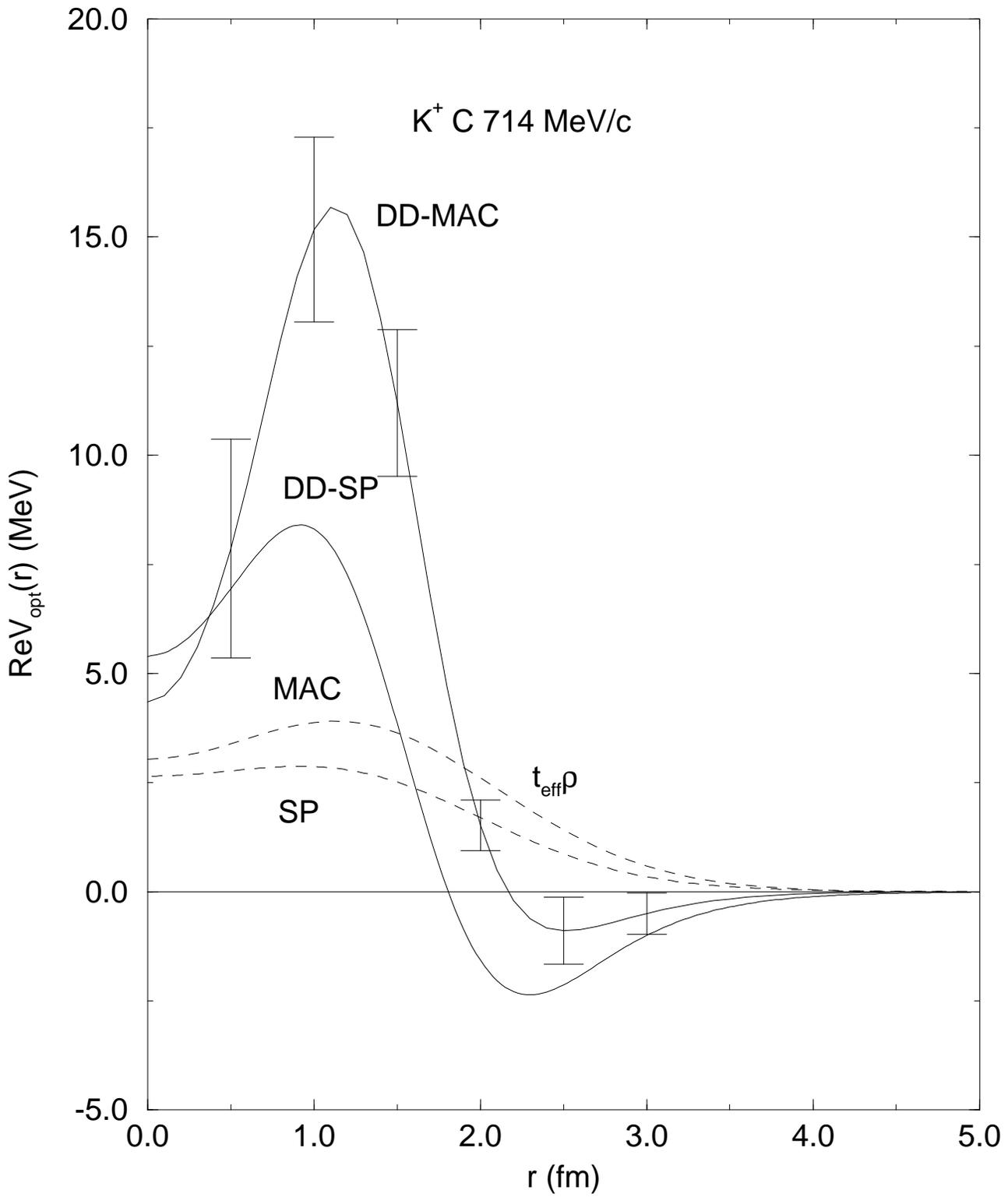
\end{document}